\documentclass[sigconf]{acmart}

\usepackage{tabularx}
\usepackage{longtable}
\usepackage{array}

\AtBeginDocument{%
  }

\newcommand{\say}[1]{\textit{``#1''}}

\copyrightyear{2025}
\acmYear{2025}
\setcopyright{rightsretained}
\acmConference[CHIWORK '25]{CHIWORK '25: Proceedings of the 4th Annual Symposium on Human-Computer Interaction for Work}{June 23--25, 2025}{Amsterdam, Netherlands}
\acmBooktitle{CHIWORK '25: Proceedings of the 4th Annual Symposium on Human-Computer Interaction for Work (CHIWORK '25), June 23--25, 2025, Amsterdam, Netherlands}
\acmPrice{}
\acmDOI{10.1145/3729176.3729200}
\acmISBN{979-8-4007-1384-2/25/06}

\begin{document}

\title{A Case Study Investigating the Role of Generative AI in Quality Evaluations of Epics in Agile Software Development}

\author{Werner Geyer}
\affiliation{%
  \institution{IBM Research}
  \city{Cambridge, MA}
  \country{USA}}
\email{werner.geyer@us.ibm.com}

\author{Jessica He}
\affiliation{%
  \institution{IBM Research}
  \city{Seattle, WA}
  \country{USA}
}
\email{jessicahe@ibm.com}

\author{Daita Sarkar}
\affiliation{%
 \institution{IBM}
 \city{Kochi}
 \country{India}
}
\email{daita.sarkar@ibm.com}

\author{Michelle Brachman}
\affiliation{%
 \institution{IBM Research}
 \city{Cambridge, MA}
 \country{USA}
}
\email{michelle.brachman@ibm.com}

\author{Chris Hammond}
\authornote{Now with Outrigger Group.}
\affiliation{%
 \institution{IBM}
 \city{Austin, TX}
 \country{USA}
}
\email{chammond@outrigger.group}

\author{Jennifer Heins}
\affiliation{%
 \institution{IBM}
 \city{Durham, NC}
 \country{USA}
}
\email{heinsj@ibm.com}

\author{Zahra Ashktorab}
\affiliation{%
 \institution{IBM Research}
 \city{Yorktown, NY}
 \country{USA}
}
\email{Zahra.Ashktorab1@ibm.com}

\author{Carlos Rosemberg}
\affiliation{%
 \institution{IBM}
 \city{}
 \country{Canada}
}
\email{carlos.rosemberg@ibm.com}

\author{Charlie Hill}
\affiliation{%
 \institution{IBM}
 \city{Cambridge, MA}
 \country{USA}
}
\email{charlesh@us.ibm.com}

\renewcommand{\shortauthors}{Geyer et al.}
\renewcommand{\shorttitle}{A Case Study Investigating the Role of Generative AI in Quality \\Evaluations of Epics in Agile Software Development}

\begin{abstract}
The broad availability of generative AI offers new opportunities to support various work domains, including agile software development. Agile epics are a key artifact for product managers to communicate requirements to stakeholders. However, in practice, they are often poorly defined, leading to churn, delivery delays, and cost overruns. In this industry case study, we investigate opportunities for large language models (LLMs) to evaluate agile epic quality in a global company. Results from a user study with 17 product managers indicate how LLM evaluations could be integrated into their work practices, including perceived values and usage in improving their epics. High levels of satisfaction indicate that agile epics are a new, viable application of AI evaluations. However, our findings also outline challenges, limitations, and adoption barriers that can inform both practitioners and researchers on the integration of such evaluations into future agile work practices.
\end{abstract}

\ccsdesc[500]{Software and its engineering~Agile software development}
\ccsdesc[500]{General and reference~Empirical studies}
\ccsdesc[500]{Human-centered computing~User studies}
\ccsdesc[500]{Computing methodologies~Natural language generation}

\keywords{Agile Software Development, Agile Epics Quality, Evaluation, Generative AI, Large Language Models, Software Requirements}

\maketitle

\section{Introduction}
Generative AI continues to be integrated into work practices to support knowledge workers' tasks with the goal of improving the quality and delivery time of their work. Large language models (LLMs) \footnote{In the remainder of this paper, we use ``generative AI'' and ``LLMs'' interchangeably.} can be effectively used to generate documents, code, or summaries, to analyze or transform data, to translate languages, to answer questions, to classify or categorize content, to brainstorm, or to plan and organize activities~\cite{cambon2023early, cardon2023generative, gao2024taxonomy, Brachman2024, wagman2025generative, wang2024surveylargelanguagemodels, zhao2024surveylargelanguagemodels}. They can also be leveraged to evaluate the quality of both AI-generated and human-created content \cite{chiang2023can, dubois2024alpacafarm, luchini2023automatic, zheng2023judging}. For example, AI evaluations have been used in education to assess student work \cite{burstein2003e, chiang2024large, song2024automated}, in the legal sector to assess fact relevance~\cite{ma2024leveraging} and LLM-generated legal texts~\cite{ryu2023retrieval}, and in software development~\cite{kumar2024llms, wadhwa2023frustrated}, where the use of AI has been an active area of research for many years (e.g. \cite{BAHI2024, Dalpiaz2021,Umar2023, Antaputra2024,Siddeshwar2024,sami2024ai}). In this work, we investigate the viability of AI evaluations in a new area of work: supporting product managers in creating high quality agile epics. Through this investigation, we contribute to the growing body of research exploring which areas of agile software development would benefit most from generative AI and how to best integrate generative AI assistance into existing workflows (e.g. \cite{cheng2024generative,Oswal2024,BAHI2024,Arora2024,lubos2024leveraging,marques2024using,norheim2024challenges}).

We chose to focus on agile epics as they play a key role in driving agile software development by aligning key stakeholders including product managers, designers, and developers. An agile epic is a representation of a body of work that can be broken down into specific tasks. Epics often describe a problem, a solution, the value to the customer, requirements (often in form of user stories), scope, and target outcomes. They represent a product deliverable that is typically too broad to be completed in a single sprint and hence are broken down into smaller, manageable tasks called user stories \cite{scrumalliance_epics, atlassian_epics}. 
Agile epics are a commonly used tool for product managers to drive work and communicate what needs to be done and why to stakeholders, as described in prior case studies~\cite{mattila2023assisting, nilsson2024optimization, olander2023agile}. High quality requirements are critical for the success of software projects \cite{Kamata2007, Hussain2016}. Poorly defined requirements in epics can result in unreliable estimation, delivery delays, churn during development, poor feature quality, and incomplete features -- all of which can lead to cost overruns \cite{cheng2024generative}. 

In this industry case study, we investigate the role that LLMs can play in evaluating agile epic quality as part of the agile software development practices of a large global company. Our goal was to understand how LLM evaluations could be effectively integrated into product managers' work practices, along with their value, usage, and limitations in improving the quality of agile epics. To this end, we conducted a user study with 17 product managers and developed two artifacts to concept test with them: (1) a rubric of quality criteria for epic elements that can be used in both AI and non-AI evaluations, and (2) an interactive, LLM-based tool called Epic Evaluator based on this rubric. Ultimately, we envision that carefully designed tooling based on LLMs will be used and integrated in future work practices to assist agile software development in effectively crafting agile epics to overcome today's challenges of poorly defined requirements. Towards our goal, we address three research questions in this case study:

\begin{itemize}
    \item RQ1: What are current epic creation work practices and potential implications for LLM-based epic evaluations in agile software development?
    \item RQ2: What are the characteristics of high quality epics that need to be considered in LLM-based evaluations?
    \item RQ3: How viable are LLM evaluations in helping product managers understand and improve agile epic quality?
\end{itemize}

This case study makes the following contributions:

\begin{itemize}
    \item We introduce and validate a rubric that can be used to assess the quality of an agile epic and outline areas for future improvement of the rubric.
    \item We introduce a tool called Epic Evaluator, including underlying prompts, that demonstrates how LLMs can be used to evaluate an agile epic.
    \item We present results from a qualitative user study and concept test with 17 product managers that provide insights into their perspectives on epic quality, opportunities and recommendations for integrating LLM evaluations into work practices, and the value, usage, and limitations of LLM evaluations. 
\end{itemize}

Our findings indicate that LLM-based epic quality evaluations are viable and desired by practitioners but need to provide flexibility to accommodate diverse practices when creating and managing agile epics. At the same time, a tool based on this approach can help organizations encourage standardization with the goal of increasing overall epic quality. 
We find strong evidence that such a tool would be useful and that there is a clear desire for support in improving epics. However, challenges remain, including the lack of domain and stakeholder knowledge when making recommendations, which affects both the quality and utility of LLM evaluations.

As a case study, our research focuses on agile software development practices in our own company. However, since agile practices are very common and similar across industries, we believe that our insights are valuable, both to researchers who want to advance LLM-assisted decision-making in areas that traditionally require more human involvement, and to practitioners who want to adopt and integrate generative AI in their organizations' future work practices.

\section{Background and Related Work}
Our paper contributes and builds upon research around software requirement creation and validation in the context of agile software development, AI for software engineering requirements and evaluation, and using large language models for evaluation.

\subsection{Agile Software Development}
Agile software development is an iterative and adaptive approach to software engineering that emphasizes collaboration, customer feedback, and continuous improvement \cite{agile_software_development}. Grounded in the principles of the Agile Manifesto \cite{agilemanifesto}, this methodology prioritizes individuals and interactions over processes and tools, functional software over extensive documentation, customer collaboration over contract negotiation, and responsiveness to change over rigid planning. While the traditional Waterfall model \cite{Kasauli2017,Royce1970} provides clear structure and predictability, it is less suited for projects with constantly evolving requirements, continuous feedback, and a need for rapid delivery and responsiveness to shifting customer needs. To overcome the limitations of Waterfall models, many companies today have adopted agile practices in their software development lifecycles \cite{Boehm2004, Dingsoyr2012}.

While the Agile Manifesto defines foundational principles, agile development practices can vary across different organizations. Agile frameworks such as Scrum \cite{Schwaber2020} and Kanban \cite{Ahmad2018} are often used to structure development into short, incremental cycles, called sprints, to adjust dynamically to evolving requirements. Work is often structured around epics, which capture a larger body of work and break it down into smaller, more manageable units called user stories, which are specific tasks or requirements that can be completed within a sprint \cite{gog_epics}. With sprints, the development process flows through short cycles typically lasting two to four weeks, during which cross-functional teams work on user stories. Agile project management software also often supports master epics, which are high-level epics that serve as containers or parent epics for multiple related epics. Compared to traditional requirement documents, epics are not fixed but evolving, do not necessarily specify all system requirements upfront, are written from a user's point of view, are lighter-weight at the documentation level, and are typically managed collaboratively by the product owner and development team. In our own company, the product development process is separated into two phases: discovery, where needs and priorities are defined, and delivery, where the final product is developed and released. Agile epics (and master epics) can be considered a tool to translate findings from the discovery phase into actionable requirements for the delivery phase of a new product or feature. 

Prior work has investigated existing requirements creation and validation processes. In agile development, there is a focus on short iteration cycles, which sometimes leads to insufficient documentation of requirements~\cite{mendes2016impacts, soares2015investigating, kasauli2021requirements}. For example, ~\citet{soares2015investigating} identified a lack of detail in user stories as one quality issue.
Research around non-functional requirements showed that they are sometimes documented formally such as in epics, but they are also documented throughout other artifacts or not at all~\cite{behutiye2020documentation, nasir2023exploratory}. One way to support improved documentation is through guidelines~\cite{behutiye2022quality}. 
Some work has begun to provide guidance on improving quality issues.
~\citet{bik2017reference} describes a method for writing user stories at a high level, including steps like validating and refining user stories. 
~\citet{ferreira2022towards} describes a set of guidelines for writing user stories, quality requirements and acceptance criteria, with suggestions like having a ``short but suggestive name.'' Our work contributes to this area of understanding by developing a rubric for agile epics, investigating users' perspectives on the rubric, and demonstrating how an LLM can leverage the rubric for automated evaluations.

\subsection{AI for Requirements Creation and Evaluation}

Prior to generative AI, tools used traditional AI to support requirements engineering~\cite{Antaputra2024, Siddeshwar2024}. A survey of practitioners showed that they still rely on prior tools, despite research advancements~\cite{Umar2023}. Yet, requirements engineering has many opportunities for generative AI augmentation, including new uses that were not previously possible with traditional AI~\cite{Umar2023, Antaputra2024, dam2019towards}.

 Research around use of generative AI for requirements engineering has focused on the early stages, like elicitation and analysis of requirements~\cite{cheng2024generative, marques2024using}. For example, \citet{Oswal2024} created a tool that helps users transform requirements into user stories using GPT-3.5. \citet{Dalpiaz2021} also developed methods to use AI to create better user stories. A case study evaluating the use of generative AI for writing epics and stories found anecdotal evidence of improved quality, team flow, and speed \cite{bockeler2024using}. They also found that domain knowledge is critical to elicit good outputs from LLMs. A preliminary investigation of the feasibility of using LLMs for requirements elicitation showed that prompts and contextual information were important, as well as experience of the users and domain knowledge~\cite{Arora2024}. A study of ChatGPT-generated requirements showed that experts often rate them as acceptable, but they can be ambiguous or infeasible~\cite{ronanki2023investigating}.

Some research has begun to investigate evaluation of epics and requirements using generative AI. \citet{Arora2024} outline the potential for LLMs to support requirements analysis, such as the evaluation of epics. \citet{sami2024ai} developed a multi-agent approach to requirements engineering using generator agents and a quality control agent. Feedback from four users indicated a potential for this kind of system to support requirements engineering work, but there is little discussion of how the evaluation criteria were created. \citet{wang2311chatcoder} designed a chat system to support refinement of requirements for software generation, focusing on precision, ambiguity and completeness to improve code generation based on requirements. \citet{lubos2024leveraging} use an LLM to evaluate software requirements based on the ISO 29148 standard \cite{ISO29148:2018}, explain the evaluations, and suggest improvements for the requirements. The LLM was able to identify many quality issues and provide valuable explanations and recommendations. Given the viability of LLM-based evaluations of software requirements demonstrated by prior work, we sought to understand its applicability to agile epics, which capture a higher level view of requirements and contain additional information to drive stakeholder alignment. Our study adds to the body of work demonstrating feasibility of the use of generative AI for requirements evaluation. Additionally, to build on work that found user expertise to be a salient factor in determining usefulness of LLM-support for traditional requirements engineering~\cite{Arora2024, ronanki2023investigating}, we further provide qualitative insight into product managers' perspectives on the utility of this kind of tool for agile software development and its potential to be used in their work practices.

\subsection{Using Large Language Models for Evaluation}
In our work, we leveraged LLMs as evaluators for agile epics. This approach, often called LLM-as-a-Judge, has become very popular in recent years to  evaluate the output of other LLMs because it allows developers to flexibly define custom criteria for their use cases without the need for ground truth. However, the same techniques can be used to evaluate text written by humans \cite{chiang2023can, dubois2024alpacafarm, luchini2023automatic, zheng2023judging, ShaerEtAl-Brainwriting-2024, GPTeach2023}, in domains such as education,  the legal sector, or software development \cite{chiang2024large, song2024automated,ma2024leveraging, ryu2023retrieval, kumar2024llms, wadhwa2023frustrated}. 
For example, Shaer et al. \cite{ShaerEtAl-Brainwriting-2024} demonstrate that GPT-4 can be used as an aid to evaluate ideas from brainstorming sessions and Markel et al. \cite{GPTeach2023} outline its use for evaluating students and teachers with GPTeach.
LLMs can perform decently well as evaluators of natural language text~\cite{zheng2023judging, dubois2024alpacafarm}, but there is also room for improvement. In expert knowledge tasks, LLM evaluators and subject matter experts do not always agree~\cite{szymanski2024limitations}. 
One way to improve performance is to fine-tune models to specifically support evaluation tasks~\cite{kim2023prometheus}.
Further tooling for the evaluator may also help.
Tools, like EvaluLLM~\cite{desmond2024evalullm} and Evallm~\cite{kim2024evallm}, enable custom evaluation of generated text using LLMs. Yet, further assistance is likely needed to support non-experts in authoring effective criteria for these kinds of tools~\cite{pan2024human}. In our work, we are not leveraging any existing LLM-as-a-Judge tools because most do not support complex, structured definitions of multiple criteria as required by our rubric. Instead, we customized prompts for Epic Evaluator as described in Section \ref{section:epic-evaluator}. Our work considers LLM evaluation in the particular domain of agile epics, contributing to the understanding of LLM usage for evaluation of domain-specific natural language text.

\section{Methodology}

\subsection{Rubric Development}
To understand existing quality criteria and how they can be applied to LLM evaluations, the first critical step in our investigation was to develop a rubric that can be used to assess agile epics. We started by reviewing top-down guidance on best practices for the software development lifecycle that have been developed and disseminated in our company. The guidelines developed by our agile development leadership team also include a template for agile epics. We used this template as a starting point to develop our rubric. The template consists of nine elements: (1) title, (2) problem statement, (3) product outcome \& instrumentation, (4) user stories, (5) requirements for each user story, (6) assumptions, (7) non-functional requirements, (8) out-of-scope, and (9) questions \& decisions. The template provided some examples for how these elements should look like but was sparse on actual quality criteria for each of the elements. For each element, we developed a definition informed by internal and external best practices and quality standards \cite{nfr, user_stories,yup_epic}, as well as input from LLMs. We used LLMs as an additional source of knowledge and inspiration, as well as an aid to articulate definitions. Based on an initial draft, we iterated with two subject matter experts (SMEs) from the internal agile development leadership team until both were satisfied with the rubric definitions. We excluded "questions \& decisions", as suggested by our SMEs, leading to a total of eight elements as shown in Table \ref{tab:abbreviated_rubric}. Detailed definitions and criteria for each element are included in Appendix~\ref{sec:rubric}). As agile epics are broader and somewhat different than traditional requirements specifications, we did not adopt ISO 29148 \cite{ISO29148:2018}. However, some of the quality characteristics from the ISO standard are similar to our rubric, such as unambiguous, verifiable, avoiding implementation details, usage of standardized structures, and representing user needs. 

\begin{table}[ht]
  \centering
  \caption{Elements of our rubric with an abbreviated definition. See Appendix  \ref{sec:rubric} for the detailed definitions and criteria.}
  \label{tab:abbreviated_rubric}
  \begin{tabular}{|p{0.2\linewidth}|p{0.7\linewidth}|}
    \hline
    \textbf{Element} & \textbf{Abbreviated Definition} \\
    
    \hline

Title & 
The title of an agile epic should be short and concise, descriptive, convey the primary objective of the epic. \\

    \hline

Problem Statement & 
The problem statement briefly summarizes the problem which is being solved in this epic including a brief description of the feature developed, and the value for the customer or user. The problem statement should be clear, comprehensive, and concise. \\

\hline

Product Outcome \& Instrumentation & 
This element describes the overall goals of this feature. Product outcomes are what determines this feature’s success. Good product goals should measure things like adoption, feature retention, satisfaction, engagement or impressions, general usage, completion percentages and speed. Instrumentation refers to metrics that can be used to measure the goals, e.g. click-throughs etc. \\

\hline

Requirements - User Stories & 
User stories describe what the user needs or expects from the system. User stories are a fundamental part of an agile epic. User stories identify and break down needs for each user role. They should be clear, structured, and complete. \\ 

\hline

Requirements - Requirements & 
The requirements element breaks down each user story into a requirement statement but typically leaves out the reason/benefit found in the user story. It is basically a breakdown of a user story into system capabilities that are required to implement the user story.\\

\hline

Assumptions & 
Assumptions are the beliefs or statements that the team holds to be true at the start of the epic but which may not yet be fully verified. They often represent things that need to be confirmed
or validated during the development process. Assumptions are typically based on what is
known at the time but can evolve or be disproved as more information becomes available. They should be explicit, testable, and describe their impact. \\

\hline

Non-Functional Requirements & 
Non-functional requirements are requirements that are not centered around the user experience. Non-functional requirements should be measurable, testable, realistic, and relevant. \\

\hline

Out-of-Scope & 
The out-of-scope section clearly states what will not be included in the feature, usually as a list of items. Out-of-scope items should be clear, relevant, specific, and justifiable.\\

\hline

  \end{tabular}
\end{table}

\subsection{Epic Evaluator: An LLM-Based Evaluation Tool}
\label{section:epic-evaluator}

\begin{figure*}
\centering
 \includegraphics[width=\textwidth]{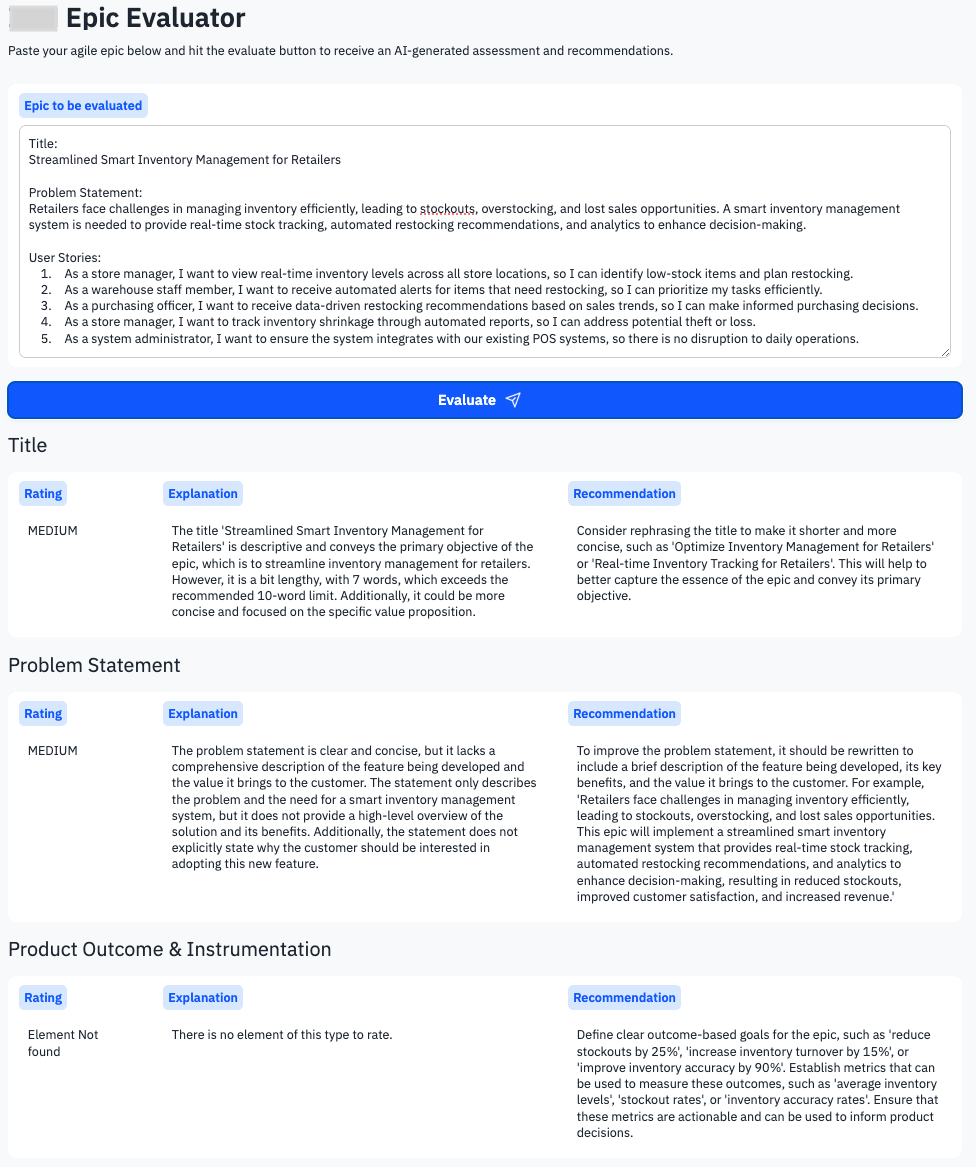}
  \caption{The experimental Epic Evaluator tool. The full text of an agile epic is pasted into the input box at the top. When pressing the 'Evaluate' button, the system checks for the presence of agile epics elements and for those present, rates the quality, provides an explanation, and makes recommendations for improvement.}
 \label{fig:epic_evaluator_tool}
\end{figure*}

Based on the rubric, we iteratively developed a number of prompts to evaluate an agile epic. We experimented with various LLMs available\footnote{Access to LLMs in our company for corporate content is limited to a set of approved models which we had to choose from.} to us (llama-3-70b-instruct, llama-3-405b-instruct, mixtral-8x7b-instruct, granite-13b-instruct) and different configurations of our prompts. After various iterations with example epics, we found that llama-3-70B-instruct yields contextually relevant and stable results with reasonable response times when used to assess each element of an agile epic separately rather than one large prompt. In our selection criteria, we defined ``contextually relevant'' evaluations as those providing examples that demonstrated an understanding of the content of the epic, and ``stable'' as those that did not change between multiple evaluations of similar epics. While llama-3-405b-instruct also yielded comparable results, response times were significantly higher. We also developed a prompt to identify the presence of each epic element in a text document. This resulted in a total number of nine prompts (see \ref{sec:appendix_prompt} for an example) that we used to develop a simple Gradio \cite{gradio} Python application, called Epic Evaluator, which takes the full text of an agile epic as input, analyses the presence of epic elements, and individually evaluates each element found (see Figure \ref{fig:epic_evaluator_tool}). Epic Evaluator leverages IBM watsonx.ai as an inference service. For each element, it generates a rating (High/Medium/Low) as defined in \ref{sec:appendix_rating_scale}, an explanation for the rating, and a recommendation on how to improve the element. While our 3-point rating scale was chosen based on our discussions with SMEs, this simple initial design was mostly intended as a probe for concept testing. The fictitious example shown in Figure \ref{fig:epic_evaluator_tool} illustrates the nature of the explanations and recommendations given. They often, but not always, contain concrete examples on how to improve an element of an epic referencing content within the epic. If an element is not found, it only generates a recommendation. We discussed this design on our team and concluded that this would be a useful initial approach that we can use for concept testing~\cite{iuso1975concept} with our study participants. This design is similar to what Lubos et al. \cite{lubos2024leveraging} studied quantitatively. In our study, we used Epic Evaluator to evaluate agile epics provided by our study participants.

\subsection{User Study}
In order to gain insights into practitioners' perspectives on epic quality, opportunities for integration of LLM evaluations into work practices, and the value, usage, and limitations of LLM evaluations, we conducted semi-structured interviews with 17 product managers. During the interviews, we used the rubric and the outputs of the Epic Evaluator tool for concept testing~\cite{iuso1975concept} -- a technique used to elicit user feedback on ideas or lightweight prototypes before investing large amounts of development resources, which was well suited for the early stages of our artifacts. Seed participants were recruited from groups of product managers who had early exposure to a technology pilot or were recommended by the agile development leadership team. We then used snowball sampling to recruit additional participants. We stopped recruiting new participants once we felt our sample was diverse enough and observations seemed to converge. The 17 participants (9 female / 8 male) were all in product management roles as Product Managers (52.9\%), Product Leads (17.6\%), and Directors (29.4\%). 47.1\% of the participants were in a people management role. Their product management experience ranged from 1.5 years to 20 years with an average of 6.6 years. 

In our interview protocol, questions were grouped into four categories: (1) background including job role and AI usage, (2) their epic creation process, importance of epic elements, and general quality considerations, (3) concept testing of the rubric we developed including rating scales, frequency of evaluation, and satisfaction, and (4) concept testing of evaluations generated by the Epic Evaluator. During section (2) of the interview, we showed participants a list of epic elements and asked them to rate their importance while thinking aloud. During section (3), we showed them the rubric and asked for feedback for each element of the rubric while thinking aloud. During section (4), we shared screenshots of evaluations of 2-4 elements of an epic they had written, which they provided to us prior to the interview, and asked them to critique the evaluations. For 5/17 participants who were not able to provide data in time for the interview, we prepared evaluations based on internal sample epics. All interviews were conducted by one moderator and one note-taker via video-conferencing and took between 45-60 minutes. Three authors conducted thematic analysis \cite{ClarkeBraun2017} of the interview transcripts by creating an initial code book based on a subset of the data. The code book was subsequently applied to all the transcripts and incrementally extended if any new themes were identified, in agreement with all three analysts. We provide descriptive statistics where appropriate but given the qualitative nature of the study, we do not provide statistical tests. The study followed company policies on user data. Participation was voluntary, with the option to withdraw anytime. Data was de-identified, stored securely, and analyzed anonymously. Participants consented to the use of de-identified data for current and future research, ensuring confidentiality.

\section{Results}
We present the themes\footnotetext[1]{https://www.mural.co/} we identified for each research question: implications of current epic creation practices for LLM evaluations, characteristics of high quality epics, and the viability of generative AI in supporting epic evaluation.

\subsection{RQ1: What are the current epic creation work practices and potential implications for LLM-based epic evaluations in agile software development?}

\subsubsection{Overview of epic creation practices}
Participants' current\footnotetext[2]{https://www.aha.io/} approaches to epic creation indicated that it is a highly \textbf{collaborative and iterative} process. Although epics are owned by product managers, who typically write the first draft, most participants noted that it evolves with input from stakeholders, which can include other product managers, lead developers, lead designers, software architects, and sometimes marketing teams. Epic creation often evolves through \textbf{multiple stages} -- as P9 explained, \say{the epic goes through different stages of being written: being committed, being in design, then ready for development, in development and ready to ship, and shipped.}

Beyond these similarities, specific \textbf{practices diverge} significantly. The starting points, tooling, formats, use of specific elements, creation time, evaluation frequency, and involvement of stakeholders at different stages can all vary. For example, on P5's team, problem identification is owned by product managers, who \say{capture it through different feedback sources}, and \say{what we need to do to solve that problem also [..] is product managers' responsibility.} In contrast, P9 approaches problem identification collaboratively: \say{we work as three-in-a-[box] and go to hear the customer pain points together, so we can capture from different aspects what they're saying.} The epic evaluation process and timing also vary: as a manager, P7 focuses on the problem statement and outcome to ensure they are aligned with the product focus, while P14 looks across elements to determine whether the right ones are present and clear. To illustrate the high variability in practices, we share two vignettes describing differing processes in Table~\ref{tab:vignettes}.

\begin{table}[htp]
    \centering
    \small
    \begin{tabularx}{\linewidth}{p{0.5cm}X}
        \hline
        \textbf{P14} &  Product managers mostly only create master epics, which they described as having a larger scope than regular epics. After receiving a request for new functionality, they work with designers and developers to ideate on solutions in Mural \footnotemark[1]. Following ideation, the product manager uses a custom Aha! \footnotemark[2] template with two structured fields: a free-form description where they enter customer personas and links to relevant artifacts (such as ideation boards and prior research), and a value statement that describes who they are solving for and what value they drive from the outcome. P14 views epics as a \say{contract with our development team}. After hand-off, developers then break down the master epic into sub-epics or issues, occasionally coming back to product managers with suggested changes for technical feasibility. \\
        \hline
        \textbf{P2} & The epic creation process starts with a new technical requirement or customer idea. They use the standard company template to write the first draft of the epic. Once the draft is ready, they work with development leads to scope and define the technical design of the feature. The product manager then writes user stories and acceptance criteria, continuing to involve developers to ensure each story is sprint-sized before handing off for implementation. \\
        \hline
    \end{tabularx}
    \caption{Vignettes of different epic creation processes}
    \Description{Two vignettes of participants' epic creation processes to show how they can vary.}
    \label{tab:vignettes}
\end{table}

Epic creation can vary within teams as well. For example, P8 writes two different types of epics. One is a \say{placeholder epic [..] to kinda get it in the system so that you can have something to note for a broader discussion,} which is held to lower quality standards and only includes a title, basic required fields, and a prioritization report. In contrast, for higher fidelity epics, they start with an \say{epic template that we've aligned on as a team} to ensure \say{uniformity} and \say{completeness.}

Another point of variability was in participants' perceived importance of different epic elements. When asked to rate the importance of each epic element from our rubric on a scale of 1-5, with 5 being most important, participants generally agreed that problem statement (M (SD) = 4.87 (0.35)) and user stories (M (SD) = 4.27 (0.86)) were most important, but there was variability in how they weighed the rest of the elements. This variability stemmed from differences in team practices and perceptions of an element's necessity. For example, P6 felt that product outcome \& instrumentation \say{is not that important because this lives higher than the epic level, at the team level,} whereas P9 found outcome very important \say{to know where we're taking the product.} Participants rated an element as less important if it is captured outside the epic (e.g., in the product strategy) or considered redundant with another element. They rated elements as more important if considered crucial for organizing, understanding, or executing the work. In some cases, participants considered an element to be theoretically important but not in practice -- for example, P13 said, \say{I would say outcome is one of the important things, but [product name] is on-prem [..] we do not have any instrumentation for seeing the adoption of the particular feature.}

\subsubsection{Opportunities for AI injection}
\label{sec:injection-opportunities}
We identified several pain points in participants' current epic creation processes that present opportunities for AI support, particularly through evaluation. 
AI-based evaluations can help teams \textbf{improve specific epic elements} that they want to work on. For example, P8 expressed a desire to improve instrumentation: \say{instrumentation, we don't do as [well] as we should, adding this to every epic, but it should be added because if it's not [..] then you lose a lot of the value in the product.} P1 noted, \say{we should probably be doing better in terms of performance requirements [..] That is definitely an area, I think, where these requirements generators can help us.} AI evaluations can also help \textbf{drive consistency}, which some participants recognized as a need either within their team or across the company. For example, P15, who manages a portfolio of multiple products, recognized that they are \say{not entirely consistent across all of [their] products} in how they scope master epics and expressed a need for \say{being a little more consistent across our portfolio in what is in a master epic.} P8 explained the importance of using a consistent format within their team for stakeholder clarity, saying that it \say{makes it easier} for designers and developers when \say{everyone is used to it.} At an organization level, P14 explained, \say{there's been pockets of people that do things differently, and all their different ways work, but either when teams cross, talk to each other, or people move from team to team [..] there's a lot of inefficiency.}
Several participants noted AI could \textbf{provide support for product managers who are overwhelmed}. P11, who works on a team in which  engineers significantly outnumber product managers, explained, \say{I think a lot of teams need help. I'm sure if you go interview engineering colleagues, they'll say PM gives us [under-specified] requirements [..] we need to do better because that's the role of PMs, is actually defining things.} They felt that \say{AI can be the accelerator} in the highly time-consuming epic creation process and spoke to the need for \say{interactive and iterative} support tools. 

Beyond the work of individual product managers, another opportunity for LLM support is in stakeholder collaboration. As the epic creation process is highly iterative and collaborative, it is prone to communication and process breakdowns, as mentioned by several participants. P14 illustrates the benefits of good epics and how they can \textbf{alleviate communication breakdowns}: \say{good epics are a huge factor in efficiency of how we’re gonna build
and deliver software that [..] if we don’t start from good epics [we are] wasting a lot of time in the margins trying to correct bad epics. So I think you’re attacking the right problem with the research here. [..] It’s probably going to save me five days worth of work in terms of meetings that I have to go into to correct that downstream [..] at a much higher expense.}

Beyond these pain points and opportunities, we also sought to identify \emph{when} LLM evaluation would be most useful in the multi-stage creation process. When we discussed how often an epic should be evaluated, a few mentioned a continuous need, such as P12 who wanted  \say{to evaluate the same epic over and over again till I know I've got it right.} However, a \textbf{pattern of common milestones} emerged as potential points of injection: (1) after completing the first draft, (2) before sharing with stakeholders, (3) after iterations with stakeholders, and (4) before starting development. The above patterns are consistent with the observation that epics are developed in stages and that elements have different levels of importance depending on the stage as P8 points out: \say{defining this up front, I think is a mistake. So again, there's a timeline aspect to this and I think that should be factored into the definition.} 

\subsubsection{Summary}
Addressing RQ1, we found that current epic creation practices offer meaningful and valuable opportunities for the injection of LLM evaluations during the creation process such as, reducing communication breakdowns, accelerating epic creation, or simply improving individual elements. As epic creation happens in stages and evaluation is needed at multiple milestones, LLM evaluation will need to take into account the current stage of an epic at the time of evaluation. Given the diversity of current practices, LLM evaluation needs to flexibly address those, but they also offer opportunities for more consistency as desired by many participants. While consistency can be achieved through standardization of what epic elements should be used or configured, standardization also comes in terms of how the quality of each epic element should be defined which brings us to our next research question.

\subsection{RQ2: What are the characteristics of high quality epics that matter for LLM-based evaluations?}
Understanding what makes an epic high quality is critical for designing LLM-based evaluations but also offers opportunities to standardize on a definition of what defines a "good" epic. In order to validate the assumptions we made when developing the rubric, we first collected participants' perspectives on epic quality without showing them the rubric, then asked them for feedback about our rubric.

\subsubsection{Product managers' perspectives on epic quality}
\label{sec:quality-aspects}
When we asked participants about their key considerations when evaluating an epic, we overwhelmingly heard \textbf{clarity and understandability} as the most important quality characteristics (9/17). The important role of epics in communicating requirements to design and development was highlighted by P5 who said \say{A good epic in my opinion, is something that you know that can get product management, development, design and architecture all clear on what we want to do, and it lists out the problem and the outcome very clearly.} It was also emphasized upon by P10 as, \say{I think the most important thing is [..] if someone from the engineering reads that epic and understands what they need to do.}

The second most mentioned characteristic was \textbf{completeness} (7/17). For some, this meant that some critical elements of an epic need to be present; for others, it meant all required elements needed to be there. P11 and P17 indicate how the lack of elements can lead to unnecessary churn or indicate communication challenges with development. As P11 explains, \say{If you don't have an acceptance criteria or high level user stories statement, that irks me. [..] I'm like, don't talk to development [..] Let's [not] waste our colleagues time [..] if you're not even trying to put some amount of acceptance criteria or any amount of critical thinking, that's when I know that epic wasn't written with any care or concern.} Similarly P17 notes, \say{If it's missing [the] trifecta of user stories, requirements and acceptance criteria.[..] just any semblance of that with complete clarity around what the expectation is for what dev's gonna build [..] That's my first sort of like OK, something's not right here.}

In addition to completeness, participants also indicated the importance of \textbf{detail and specificity} (6/17). As a measure, some participants indicated that an epic should be understandable to a newcomer or peer. As P14 puts it, \say{Does it have all of the right components documented? Is there enough detail in it? Can I understand the story? And you might test that by maybe asking a peer product manager who hasn't been involved in the process if they could read it and they could kind of understand it.} On the other hand, P6 was concerned about excessive content, stating,\say{One is definitely how how easy to read it is. I think it's really easy [..] as a PM to make your epics like unnecessarily lengthy and wordy because the the main thing that you're doing is writing the epic. And so you put a lot of detail into it, but then you take it to the devs and designers [..] and it's [..] very hard [..] to digest [the] information.}

Several participants highlighted "semantic" quality characteristics related to the content of an epic, such as meeting customer needs, driving value, or alignment with strategy, which may require contextual knowledge beyond the epic itself and could be more challenging for an LLM to evaluate. As P4 explains, \say{One other field that I think is important is the customer value statement [..] Like what is the end value to a customer once we implement this particular functionality?} P7 adds, \say{I'm first looking at alignment [..] so instead of reading the whole [..] thing, I'm gonna look at the problem statement, and then I'm gonna look at the outcome and whether this is aligned to where my focus is [..] I noticed that we were doing a lot of stuff that was not tied to a coherent strategy.}

One manager (P8) offered his own perspective on how he would go about evaluating the quality of epics of his reports: \say{So the first thing I'm evaluating as a leader is (1) Is this aligned to what we're focused on and if not, why? [..] (2) What the requirements are, so that I'm now going to the user stories [..] based on the problem that we're trying to solve, like do I actually think that we understand what the user is struggling with and what we're helping them with [..] if that jives, then I'm looking at the completeness of everything else.}

Other quality aspects mentioned less frequently during the interviews were: following structural guidelines, including links to outside documentation, test cases, and dependencies, ease-of-adoption by adjacent services, restricting the epic to the "what" and not "how" it is being implemented, and the need for epic elements that were not part of the standard company template.

\subsubsection{Product managers' feedback on the rubric}
When we showed participants the rubric, the majority of them  expressed their satisfaction (12/17). For instance, P1 said, \say{Overall it's very good. It's actually kind of nice to have you guys kind of looking at it this way, [..] more of a scientific view on something that was maybe more of an artwork writing requirements, right.}  with some calling it comprehensive, concise, clear, and even scientific, and many were interested in using it as is as a guideline for epic quality. 

From their reactions, we had the impression that the level of detail was unexpected but not unwelcome, which was surprising given the diversity of epic creation practices outlined in RQ1. While many were not strictly following the "official" guidelines, participants were mostly familiar with the elements in the rubric, minus some confusion about specific elements (6/17). Most welcomed the clear definitions of how these elements should be written as encouragement for them to strive towards better epics for themselves or their teams. As P14 mentioned, \say{I absolutely would use this rubric. I would say in reality today only some of it right. Some of it is intuitive. Yep, that's exactly [what we are] already doing [..]. There's other pieces that I was like, yeah, definitely. I would love to have a check and balance on that because I know it's something that slips through the cracks on me.} P16 humorously added, \say{As a user, I would not use any sort of an AI tool to evaluate it after the fact, I would use it to actually write it in the 1st place as a leader.} 

As part of the think-aloud exercise when concept-testing the rubric, we identified several pain points with the rubric that describe additional needs when evaluating an epic.  

Several participants pointed to \textbf{missing elements} required to support their practices. Most notably, the standard template we used for the rubric did not include acceptance criteria for user stories, which many participants (7/17) mentioned to be critical for their epics. In RQ1, we discovered that the problem statement appears to be the most important element for a rubric, and most agreed with the definition of the quality for this element. However, multiple participants felt the problem statement should only focus on the actual problem, whereas solution and customer value should be broken out as new elements, indicating that the element may be overloaded.

The variability we discovered in RQ1 continued to reflect an important need for participants, as some expressed that the rubric may be too \textbf{rigid} for them in terms of how quality is defined. For example, P2 commented on their team's selective use of epic elements: \say{everything in this table, like non-functional requirements, out-of-scope, and assumptions -- this is something that [..] sometimes we will ignore it, because it's not always relevant. So I guess I wouldn't want to have a low score because [..] this section is not perfect.} Participants did not want to be assessed negatively if using an element differently in their practice, or if not using it, pointing to the need for a customizable rubric.

The \textbf{inconsistent use of epics and master epics} came up as another challenge for measuring the quality of an epic based on a rubric. Some participants mentioned that some elements in our rubric, such as product outcome and instrumentation or assumptions, are defined at a higher level in master epics or product strategy documents; other participants only use master epics as the structure to represent an epic. Dependencies of elements across the two constructs should be taken into consideration when evaluating an epic or master epic so that product managers do not get penalized for missing elements.

Several participants  mentioned the importance of \textbf{edge or boundary cases} in user stories and out-of-scope, as explained by P4 \say{clarify[s] that this needs to be implemented or this doesn't need to be implemented or this needs to be tested or this doesn't need to be tested}. Our rubric did not specifically emphasize those cases, and evaluations and recommendations for them will require a deeper semantic understanding of the feature and its domain versus simply testing attributes such as clarity and conciseness.

Finally, several participants pointed out that our rubric was \textbf{lacking guidance on sizing}, an important factor because epics should not be too small in scope, typically larger than what can get accomplished in a sprint, but also not too large. This scope of an epic is typically reflected in user stories and requirements but the number of stories and requirements are not a good predictor for the scope necessarily. Proper sizing requires experience and deep product knowledge, which could pose a challenge for an LLM-based evaluation.

\subsubsection{Summary}
For RQ2, our findings indicate that product managers are mostly satisfied with the rubric we developed but pointed out several pain points that need to be addressed, such as missing elements, rigidity, testing for edge and boundary cases, lack of sizing, or getting penalized for missing elements because of the structural breakdown into master epics and epics. The rubric is comprehensive and covers many of the quality criteria participants mentioned to us before showing them the rubric; however, evaluating semantic quality aspects of an epic (strategy alignment, sizing, customer needs) requires product or domain knowledge which poses a challenge for LLMs. It is also interesting to note that "completeness" was mentioned as a quality criteria while at the same time, participants did not want to get penalized for missing elements. Consistent with our findings in RQ1, this suggests that LLM evaluation needs to be stage-aware and customizable to stage and team-specific definitions of "complete."

\subsection{RQ3: How viable is generative AI for supporting product managers in evaluating epics?}

\subsubsection{Current uses and perceptions of AI}
\label{sec:current-AI}

The majority of the participants reported using generative AI in some capacity in their daily work (15/17). Uses of generative AI among participants mostly included some form of \textbf{content generation and summarisation}. In the context of agile epics, it is mostly used during discovery for \textbf{market research} and during epic creation to write \textbf{requirements, user stories, and test scenarios}.

Nearly half of the participants used an internal experimental tool for this purpose, as described by P12: \say{Once I've kind of got a structure and subject matter content together, then I use AI for writing it, paraphrasing it, condensing it, summarising it, et cetera.}
The 2 participants who were not using AI cited either a \textbf{lack of need or a lack of awareness of suitable internal tools} as the primary reason. 

When asked about their confidence in AI-generated output, most participants expressed a high level of \textbf{trust for low-risk, simple and structured tasks}, viewing AI-generated content as a valuable starting point. P6 noted, \say{I mean it's all relatively low risk, things that I'm doing. It's not like I'm having it interact with any systems, and ultimately I can edit the content as much as I want before I do anything else with it. So I feel totally fine using it now.} However, nearly all participants emphasized the importance of a \textbf{human-in-the-loop} approach, where they retain control to edit and refine the AI output. For example, P5 shared their envisioned use of the tool: \say{we take what the AI tools generate as a starter and add our own details and elements to it.}

\subsubsection{Values and desired uses of LLM evaluations}
In reviewing the Epic Evaluator's assessment of their sample epics, 15/17 participants agreed with aspects of the score or recommendations. Similar to their positive sentiments on the rubric, some participants expressed enthusiasm for adopting a tool like Epic Evaluator: P14 said the evaluation is \say{really good [..] I really want to use it right now,} P2 compared it to\say{my peer giving me feedback,} and P8 said, \say{I'm very excited about this [..] I think this will be a great tool.}

Participants' enthusiasm stemmed in part from identifying several specific uses in which LLM evaluations can provide value, indicating that generative AI can be viable in supporting epic evaluation given sufficient oversight and control over outcomes. Many participants indicated that they would use an LLM-based evaluation in their work to \textbf{augment stakeholder and peer feedback} (11/17) -- as P15 explained, an LLM evaluation is \say{not in lieu of an experienced product manager and getting their feedback, but it gives them another tool to build strong [..] epics.} Participants envisioned using LLM evaluations as a \textbf{starting point}, to drive \textbf{general improvements}, to get \textbf{accelerated feedback} or to evaluate and improve a \textbf{specific aspect of epic quality}, particularly those described in Section~\ref{sec:quality-aspects}. For example, P5 wanted to assess clarity: \say{I would use it to make sure [..] we provide everything clear and crisp for the development and design to take place.} Leveraging the strengths of automated evaluation, some
participants felt that the tool would be useful for \textbf{bulk evaluations} of epics across their team and identifying opportunities for improvement. P8, a team manager, commented on the difficulty of managing large quantities of epics within their team and the potential for LLM support: \say{We have a lot of them [..] at least two pages worth of master epics in here. So if I could, for example, [..] be like, analyze all the master epics in this report, give me feedback, [..] I would definitely try that.} Other team managers also envisioned \textbf{team usage} of LLM evaluations; some
specifically felt that it could act as a \textbf{training tool} to help novice product managers write higher quality epics. However, some 
managers also specified that they would not use this tool to evaluate employee performance, and one non-manager similarly expressed apprehension: \say{[if we're] doing the work just for the sake of the score, I'll be starting to worry - isn't my manager being reported on the score?}

\subsubsection{Needs and barriers for LLM evaluation}
To understand \emph{how} an LLM should provide feedback, we reviewed participants' responses on their preferred rating systems. Participants shared mixed opinions on the idea of receiving ratings and how they should be given, but largely agreed that \textbf{clear, actionable recommendations should be the focus}, as this type of feedback would help them improve their epics. As P2 explained, \say{verbal feedback is more actionable than scores,} and \say{for me, the important thing will be, what do I need to improve.} Some expressed reservations about scoring in general, including the \textbf{fear of being ranked, uncertainty and inconsistencies} in score interpretations, and the potential for scores to distract from the primary goal of improving epic quality. For example, P5 said, \say{if we introduce numbers, some people would always want to have a high score or, you know, 100 out of 100, and that might introduce iterations that might cause delays.} On the other hand, some felt that including a score to accompany recommendations was acceptable or even desirable for motivating them to make improvements. Among those who supported the idea of scoring, there was a preference for a multi-parameter approach. Suggestions included rating individual aspects of the epic, such as clarity and completeness, rather than assigning a single overall score. Many participants suggested using a \textbf{three-point scale (e.g., high-medium-low)} to avoid \say{splitting hairs} over more granular scales (7/17). However, a three-point scale may be less useful if the majority of scores land in the middle, as some of our participants experienced. P7 commented, \say{I didn't get anything because both of my things were middle of the road, [..] who cares, you know? There's a lot of things that could be 5 out of 10. So I would have liked it to emphasize on one side of the spectrum.}

In discussing instances of disagreement with the LLM evaluation, participants also raised barriers to adoption and suggestions to improve the value provided by LLM evaluations. Several participants wanted more \textbf{specificity}, either to content in the epic or to their domain (6/17). P11 gave an example of desired specificity: \say{especially for NFRs, which might tend to be a long list, is there a way to kind of say, these specific ones need work?} Participants also wanted more \textbf{actionable} recommendations - ones that they can implement to meaningfully improve epic quality (5/17). In some cases, recommendations were less actionable due to \textbf{lack of feasibility}. As P9 explained, \say{many of the things are dependent on information I have no access [..] so as much as I agree with the points, I would have not been able to change it much after reading the recommendations anyway.}

Among the 17 participants, 5 discussed the need for LLMs to have \textbf{domain and stakeholder knowledge} to improve specificity and actionability. P12 gave an example of how the LLM's lack of stakeholder awareness resulted in unhelpful recommendations: \say{for my product line, when I talk in terms of post-purchase fulfillment organization as a keyword, my [stakeholders..], they all get it. [..] A tool that is untrained will not be able to make sense of it and will keep asking for more and more information.} P7 called out a similar instance in which the LLM incorrectly flagged their product name as jargon: \say{[The stakeholders] are assigned to the product, they know what [product name] is, [..] so I don't know that I would classify the product name for my product as [..] jargon.} Some participants questioned the LLM's ability to assess the quality of epic elements without domain awareness. As P17 noted, \say{what's hard for outcomes and instrumentation is the number to measure with. I don't know if the system knows that this thing is even measurable.}

Another major cause of poor feedback was the LLM's lack of consideration for variability in epic creation practices (5/17). Some participants were concerned being penalized for differing formats, as highlighted by P6, \say{to me, you can have an acceptable user story that doesn't necessarily so rigidly follow the, 'as a user, I want [goal] so that [benefit]' format}. Others use epic elements in unique ways that were not accounted for in the development of Epic Evaluator, such as P2, who said, \say{the way we usually use the nonfunctional requirements is for me, as a PM, [..] I am the only one who will read what I write in this section, [..] so this is why the feedback here, it's not very actionable. Is it important that it's not clear if it's something that I am writing for myself? Probably not.} Others were concerned about being penalized for missing elements, either during earlier stages of creation or for elements not applicable to their team. In these cases, they suggested focusing on evaluating individual elements rather than the overall epic.
Participants also noted that epics can vary in type and complexity: as P8 recommended to \say{first, categorize the type of epic and then depending on the type of epic, apply a specific rating scale to it.} Due to the Epic Evaluator's current limitation in evaluating diverse epics, elements, and formats, P2 said, \say{the concern would be about [..] ranking epics based on this because I guess every team has like different methodologies.} Hence, participants felt that LLM feedback should \textbf{accommodate variability in practices} and avoid penalties for discrepancies.

Given these limitations, participants emphasized the need for human-in-the-loop when using LLM-based epic evaluations, similar to their approach to existing AI tools (see~\ref{sec:current-AI}). 

As P12 explained, \say{I'm not able to trust any assistant completely. I will want to review it. I will read a recommendation it gives back to me and apply my own judgement to decide whether I can trust it completely.} They felt that they could trust the evaluation to some degree due to its similarity to human feedback and its usefulness -- P10 felt that \say{the recommendation is really good [..] and again, it's up to me if I find it not applicable, maybe I will not apply the recommendation, but it is still very helpful.} On the other hand, participants also recognized limitations in LLMs' contextual knowledge that lowered their trust. For example, P17 questioned, \say{is it really giving me the right recommendation? [..] It's not trained on my product and my data, so it doesn't really know that I have another epic that's doing something else related to this.} Trust was also lower when participants themselves lacked knowledge of the domain and hence could not properly assess the quality of LLM outputs. Participants' ratings of trust in the LLM evaluation reflected these mixed sentiments. Among the participants who rated their trust on a scale of 1-5, with 5 indicating highest trust, the average rating was 3.5. As P17 summarized, \say{I think complete trust would mean it makes blind changes without my involvement, and it's not that.}

Finally, participants discussed the importance of \textbf{integration into their existing workflows} to ensure LLM evaluations do not create additional overhead. P6 cited this consideration as a major potential barrier to adoption: \say{the friction of going into Aha, copying [the epic], going into the chat, pasting it like that back and forth, that is probably the main element that would stop me.} Similarly, P14 said, \say{meeting me in the tools where I'm working is probably the biggest requirement for me.}

\subsubsection{Summary} To address RQ3, we found that most of the product managers we interviewed were already using generative AI in their day-to-day work which is a great basis for the adoption of LLM evaluations. Participants were enthusiastic about Epic Evaluator and saw value in specific uses such as augmenting stakeholder and peer feedback, general improvements, accelerated feedback, bulk evaluations for managers, and as a training tool for novices. We observed additional needs and potential barriers to adoption including lack of specificity, need for actionable recommendations, lack of domain and stakeholder knowledge, importance of integration into their existing workflows, and lack of consideration for variability in their practices (consistent with findings in RQ2). Ratings seemed less important as compared to actionable recommendations, and a three-point scale appeared to be sufficient for most. In the following section, we share trade-off considerations and recommendations to improve the value of LLM evaluations.

\section{Discussion and Implications}
This work sought to investigate the use of AI evaluations in a new application area: agile epics. Agile epics are a critical artifact in the software development life cycle as they communicate detailed requirements to design and development. Creating high quality epics is a time-consuming task for product managers. Nonetheless, many of our participants established practices to improve the quality of epics for themselves or for their teams through peer reviews, manual scoring systems used for training new product managers, developing their own quality standards for their teams, and welcoming more standardization across the company. Hence, leveraging generative AI to help with quality assurance was an exciting prospect for our participants. As such, LLM evaluations 
can help time-pressured product managers craft higher quality epics, alleviating downstream problems such as churn, communication breakdowns, product delays, and cost overruns.

Our findings indicate that LLM evaluations of epics are viable and can provide value today. Participants were most appreciative of the explanations and recommendations that contained concrete examples from the epic. They were less interested in getting rated but rather being advised on how to improve their epics more efficiently. As such, they expressed strong interest in actionable suggestions, which blurs the boundaries between an epic evaluation and authoring tool. Managers saw value in knowing which epics from their team need their attention and an easy way to spot them among a larger body of epics.

\subsection{LLM evaluations should be augmented with domain-specific knowledge}
While explanations and recommendations were often grounded in the epic being evaluated, sometimes they didn't make sense because of the lack of domain or context knowledge. Domain knowledge includes, for example, product information, product-specific terms, related epics, overall objectives from master epics, and stakeholder information. The lack of domain knowledge has implications for properly measuring the "semantic" qualities of epics but also sometimes leads to a lack of specificity or actionable guidance. 

This problem can be addressed through both human-driven and automated approaches. For example, product managers could augment LLM evaluations with domain-specific knowledge relevant to the evaluation task. However, asking product managers to manually add this knowledge would introduce additional barriers. A retrieval augmented generation (RAG) approach may be more feasible \cite{lewis2020retrieval}. However, relevant content would need to be curated and stored in a knowledge base. As much of the background research for a new product often happens during the discovery phase, we believe an interesting approach would be to make that knowledge available through vector databases or search indices \cite{Elasticsearch2025, Milvus2025} so that it can be accessed through a RAG system during epic evaluations.

\subsection{Ensure ratings are clear and actionable}
While ratings were perceived as less important, they remain necessary for drawing attention to parts that require improvement, particularly for managers who wanted to use LLMs for bulk evaluations. Their lower perceived importance could be due to the scores participants received in the study.
As some of our participants observed, the medium rating was most prevalent in the samples we evaluated. 
This may indicate an issue with the definition of our rating scale, and it validates previously identified challenges that LLM evaluators often face in deciding between multiple options, which can be measured through positional bias \cite{ashktorab2024aligninghumanllmjudgments, kim2023prometheus}. Findings are inconclusive on what the best approach is, but they point to users' needs for clear, distinct ratings as a characteristic of actionable LLM evaluations.
To address these needs, we aim to further experiment with rating scales that have more differentiating definitions or point systems that can more clearly correlate between high and low epic quality. Inspired by some more recent LLM-as-a-Judge approaches \cite{wagner2024blackboxuncertaintyquantificationmethod}, 
a Chain-of-Thought approach \cite{Wei2022} could be used by first asking an LLM to recommend improvements or produce a general assessment of an element, then feeding this output into a subsequent evaluation step with a rating scale defined based on the types of recommendations.

\subsection{LLM evaluations should be stage-aware}

When we discussed the frequency of evaluation with our participants, a pattern for injecting LLM evaluations into their work practices evolved. While one asked for real-time evaluations, the majority seemed to converge around four stages in the creation process: (1) after finishing an initial draft, (2) after refining the draft with other product managers on the team, and before sharing with design and development, (3) after discussions with design and development leads, and (4) before delivering to design and development. However, our interviews revealed that participants were concerned about an evaluation tool that gives non-favorable or less useful ratings to parts of an epic that are not well developed yet because, as many of them emphasized, an epic goes through stages. An epic often starts with title, problem statement and user stories, and other elements are added and evolve over time in a collaborative process. While many mentioned completeness as an important quality criteria, epics in earlier stages are not necessarily low quality because they are not complete. An evaluation tool should be aware of the current stage of an epic, provide an option for users to indicate the stage, or allow users to select which elements to evaluate at a given time. Stage-aware evaluations also have implications for generating an overall epic rating, since elements that are not relevant at a certain stage should not be included in the overall score.  

\subsection{LLM support should be tightly integrated into existing agile management tools}

While many participants reacted positively about using an LLM evaluation tool, they indicated that it needs to be tightly integrated into their existing epic management applications so it's easy-to-use on demand. While our experimental Epic Evaluator tool provides simple way to evaluate epics through pasting the text of the full epic into a text box, it requires users to extract the elements of an epic beforehand. This poses a challenge given the diversity of tools being used in our company and the somewhat limited export capabilities. We experienced this first-hand when we converted epics that participants shared with us to evaluate for our concept testing. For tooling to be adopted, there needs to be either an integration at the API level or it needs to be integrated as a plugin into existing agile management tools. For future versions of our tool,  we are currently exploring various design integrations. 

\subsection{LLM evaluations should drive consistency in epics while providing flexibility and user control to support diverse practices}

We were surprised by the diversity of existing epic creation practices across our organization despite the existence of a standard template. We learned that different teams had created their own templates based on or inspired by the standard template, while other teams had developed their own methodology from the ground up or used approaches that came in through company acquisitions. 
Despite diversity in practices, the standard elements we used for our rubric seemed familiar to most, and more standardization was viewed as acceptable and even desirable. There mostly seemed to be an agreement that problem statement, and user stories (including requirements) are the most important elements of an epic. Importance seemed less clear with other elements. Overall, the rubric and the quality criteria we had developed seemed valuable to participants, but the rubric was missing some elements they used in their practices, and we received feedback on how to improve the quality criteria in various ways. While some improvements may require more research, such as the addition of domain-specific knowledge as outlined above, other changes will be easily incorporated into the next version of the rubric. In order to support the need and desire for standardization while preserving some flexibility, we propose the creation of a core template and extensions that support different practices. The rubric would be aligned and customizable to the teams' needs at different stages, and an evaluation tool should be built on a custom rubric.

\subsection{Beyond LLM support}
Ultimately, we recognize that LLMs are only one piece of the puzzle in addressing product managers' needs, with the potential to augment parts of their work but not to automate it entirely. One way to look at agile epics is that they serve as an instrument to translate findings from the discovery phase into actionable requirements for the delivery phase of a new product or feature through a highly collaborative, iterative process. While our findings indicate that LLM evaluations of epics can improve aspects of this process, a holistic approach is needed to fully address communication breakdowns from discovery to delivery.
Participants pointed out other causes of breakdowns, such as the usage of diverse and disconnected tools and limited stakeholder access to these tools. While our standard template contains a section on questions and decisions meant to encourage stakeholder discussions around an epic, those discussions, in practice, happen mostly outside the main agile management tool using Slack, email, file sharing, or other online discussion forums. In addition to leveraging LLM evaluation tools, we also suggest practitioners review their existing agile project management tooling portfolio to further reduce communication breakdowns. As such, product managers remain indispensable in the complex work of managing epics and cross-functional coordination.

\section{Limitations}
As an industry case study, our research was conducted in our own company, potentially limiting our findings. First, we were able to experiment only with LLMs that were approved for safe use within IBM. This excluded the use of closed source models such as, GPT-4 in our model evaluation -- those models may outperform the models we used. Second, while we have iterated over our prompts, we acknowledge that prompts are known to be sensitive to even small changes and changes of the LLM being used and may not yield the same performance under different conditions \cite{rottger-etal-2024-political}. Third, although agile practices have similarities across many companies, they may not be the same. We acknowledge that our research may not be readily transferrable to other organizations because of IBM-specific agile practices, for example, related to elements of an epic being used, epic creation process and stakeholders, or quality expectations. Last, since we concept-tested LLM output through screenshots, our participants did not have the opportunity to interact directly with the tool we created. While this approach elicited valuable insight into the viability and barriers of our rubric and LLM-assisted epic evaluations, it may limit insights into real-world usage.

\section{Conclusion}
This industry case study explored the potential of using LLMs to evaluate the quality of agile epics, a core component of agile software development. Interviews with 17 product managers of varying experience levels, during which we also concept-tested a rubric and LLM evaluations based on the rubric, highlighted both the opportunities and limitations of LLM evaluations in supporting the epic creation process in our company. Our participants were enthusiastic about using LLM evaluations to improve the quality of their epics or their team's epics, addressing problems of churn, communication breakdowns, product delays, and cost overruns. At the same time, our findings also underscore the need for refinement to make this approach more effective and adaptable to existing agile workflows. Our future work focuses on improving explanations and recommendations through the integration of domain knowledge as a fundamental challenge. We also plan to extend this research to incorporate views of other stakeholders, such as design and development, who are on the receiving end of an agile epic.

While this study was in the context of agile practices in our company, we believe that lessons learned from this work are meaningful for both the "reflective practitioner" \cite{Schon83} who seeks to infuse generative AI capabilities into agile work practices in their own organization and the researcher exploring areas of AI usage that have been traditionally very challenging as they require a lot of human decision-making, insight, and oversight. In practice, achieving performance improvements by integrating AI assistance into human workflows is challenging and may even lead to performance degradation \cite{campero2022testevaluatingperformancehumancomputer}. AI assistance needs to be carefully designed to be successful, and we believe that our research has multiple contributions towards that goal, including a novel rubric that we validated, a set of prompts for quality evaluation, detailed insights into current agile practices, the role of LLM evaluations in quality evaluations, and validation that LLM evaluations are a viable approach to facilitate how agile software teams create and manage epics in the future to successfully deliver high-quality, user-centered software products.

\begin{acks}
We would like to express our sincere gratitude to the members of the agile software development leadership team and the AI productivity team at IBM for their invaluable support and collaboration throughout this project.

A special thanks to Ed Jacob-Moffatt, Gord Davison, Michael Desmond, Praveen Anthony, and Adam Chase for their guidance, constructive feedback, and continuous encouragement. Their expertise and commitment have significantly contributed to the development and refinement of our ideas. 
\end{acks}

\bibliographystyle{ACM-Reference-Format}
\bibliography{main.bbl}

\newpage
\appendix

\onecolumn
\section{Appendix}

\subsection{Rubric}
\label{sec:rubric}

\begin{longtable}{|p{2cm}|p{12cm}|}
\caption{The rubric developed with subject matter experts. The rubric includes eight elements. For each element, we provide a definition and criteria as well as an example.} 
\label{tab:paginated-table} 
\\ \hline

\textbf{Element} & \textbf{Definition} \\ \hline
\endfirsthead
\hline
\textbf{Element} & \textbf{Definition} \\ \hline
\endhead
\hline
\endfoot

Title & 
The title of an agile epic should be:

\begin{itemize}
\item Short and concise (no more than 10 words),
\item Descriptive (capture the essence of the epic), and
\item Convey the primary objective of the epic (value of doing the epic).
\end{itemize}

Example: “Enable Password Reset for Registered Users to Regain Account Access.”
\\ \hline

Problem Statement & 
The problem statement briefly summarizes the problem which is being solved in this epic. It also includes a brief description of the feature developed that addresses the problem, and the value for the customer or user and why they should be interested in adopting this new feature.

The problem statement should be:

\begin{itemize}
    \item Clear: Free of jargon and ambiguity, the description should be easy for all team members and stakeholders to understand.
    \item Comprehensive: The problem statement needs to contain three elements: (1) A description of the problem being solved, (2) a high level description of the feature addressing , and (3) the value for the customer.
    \item Concise: It should be written without unnecessary detail or redundancy.
\end{itemize}

Example: “Users struggle to regain access to their accounts when they forget their passwords. The current process is difficult and customers are abandoning the product or require help from support agents. This epic will implement a password reset feature, integrated with existing user management systems, to enable users to regain access to their account. This feature will reduce customer support requests for password recovery by 30\%, improving user retention and reducing operational costs.”
\\ \hline

Product Outcome \& Instrumentation & 

The “product outcome and instrumentations” element describes the overall goals of this feature. Product outcomes are what determines this feature’s success. Good product goals should measure things like adoption, feature retention, satisfaction, engagement or impressions, general usage, completion percentages and speed. Bad product goals are measurements around lagging metrics like churn, revenue, customer health, renewal, etc. Instrumentation refers to metrics that can be used to measure the goals, e.g. click-throughs etc.

Example: “Outcome: Reduce time to recover access by 50\%, or increase number of users that can recover access to their account without help from support, Metric: Reduced number of support tickets on this topic”
\\ \hline

Requirements 
- User Stories & 

User stories describe what the user needs or expects from the system. User stories are a fundamental part of an agile epic. They should be:

\begin{itemize}
    \item Clear: They should be written in a clear and user-centric way to communicate requirements.
    \item Structured; They should be written in a structured and consistent format to ensure clarity, understanding, and alignment across the team.
    \item Complete: They should also be complete in terms of what is needed for the overall feature. If there are too many user stories, the epic may need to be broken down.
\end{itemize}

The structure should be as follows:
As a [user role], I want [goal] so that [reason/benefit].

The who [user role], what [goal], why [benefit] format avoids getting into the how (as that should not be part of the epic).

User stories identify and breakdown needs for each user role. If necessary, they can summarize bigger needs and then break down them down into sub-details. These details define the epic / feature needs.

Example: “As a registered user, I need to be able to reset my password so that I can regain access to my account if I forget it.”

\\ \hline
Requirements 
- Requirements & 

The requirements element breaks down each user story into a requirement statement but typically leaves out the reason/benefit found in the user story. It is basically a breakdown of a user story into system capabilities that are required to implement the user story.

Example for the above user story:

“The registered user should be able to access a password reset function”
“The registered user will need to be validated prior to executing the reset”
“The registered user should be able to choose a new password”

\\ \hline

Assumptions & 

Assumptions are the beliefs or statements that the team holds to be true at the start of the epic but which may not yet be fully verified. They often represent things that need to be confirmed or validated during the development process. Assumptions are typically based on what is known at the time but can evolve or be disproved as more information becomes available.

\begin{itemize}
    \item Technical Assumption: "The current API version will support the new feature without requiring significant changes."
    \item Business Assumption: "Users will find the new filtering functionality useful and increase their product search efficiency."
    \item User Behavior Assumption: "The majority of users will use this feature on mobile devices rather than desktop."
\end{itemize}

Assumptions provide guardrails for the feature being developed. They should be:
\begin{itemize}
    \item Explicit: Assumptions should be clearly stated and not left to implicit understanding.
    \item Testable: Where possible, assumptions should be validated through early testing or research.
    \item Impact: Their impact should be explained when listed.

Example: “We assume that the existing user database can handle additional password reset requests without performance issues. This will be validated during testing.”
\end{itemize}

\\ \hline

Non-Functional Requirements & 

Non-Functional requirements are requirements that are not centered around the user experience. Examples include how development needs to handle cookie management, or how long data should be held for, or what sort of rate limits should be expected. Non-functional requirements should be:

\begin{itemize}
    \item Measurable: Non-functional requirements should be quantifiable (e.g., performance, security, usability).
    \item Testable: Non-functional requirements should have clear criteria for validation.
    \item Realistic: The requirements should be achievable within the project's constraints and technology stack.
    \item Relevant: Non-functional requirements should address critical aspects like performance, security, or scalability, and not add unnecessary overhead.
\end{itemize}

Example: “The system should respond to password reset requests within 1 second for 95

\\ \hline

Out-of-Scope & 
The out-of-scope section clearly states what will not be included in the feature, usually as a list of items. Out-of-scope items should be:

\begin{itemize}
    \item Clear: Out-of-scope items should be clearly and explicitly defined, leaving no ambiguity. Stakeholders and team members should understand exactly what is being excluded.
    \item Relevant: The out-of-scope items should be directly related to the epic, so they define realistic boundaries. For example, listing "completely different features" (like reporting or analytics) might be irrelevant if the epic is about search functionality.
    \item Specific: High-quality out-of-scope items should specify particular features, functionalities, or tasks that will not be included. Vague statements (e.g., "some advanced features") can lead to confusion.
    \item Justifiable: There should be a clear rationale for excluding certain elements. It could be due to time constraints, resource limitations, or the need to prioritize other features first.
\end{itemize}

Example: “Out of Scope: Two-factor authentication (2FA) will not be implemented in this release.”

\\ \hline

\end{longtable}
\twocolumn

\subsection{Rating Scale Used}
\label{sec:appendix_rating_scale}

\begin{verbatim}
"""
- HIGH: HIGH quality means almost all requirements in the 
QUALITY STANDARDS FOR {element_name} elements are met. 
The quality is very good and needs little to no improvements.
- MEDIUM: MEDIUM quality means some requirements in the 
QUALITY STANDARDS FOR {element_name} elements are met. 
The quality is generally good but can be improved.
- LOW: LOW quality means many of the requirements in the 
QUALITY STANDARDS FOR {element_name} elements are not met. 
The quality is bad and needs improvement.
"""    
\end{verbatim}

\subsection{Example Evaluation Prompt for Title Element}
\label{sec:appendix_prompt}
\begin{verbatim}
"""
You are a product manager and subject matter expert in 
agile epics. Your task is to assess the AGILE EPIC below. 

AGILE EPIC
{epic}

For this ASSESSMENT, please rate the following element 
of the agile epic: TITLE

In your ASSESSMENT rate the QUALITY of the TITLE on the 
following scale:

{scale}

Please rovide a detailed EXPLANATION for your assessment 
with examples where the quality suffers and make 
RECOMMENDATIONS on how to improve the TITLE. Follow the 
QUALITY STANDARDS FOR TITLE strictly. 

QUALITY STANDARDS FOR TITLE elements:
The title of an agile epic should be: 
- Short and concise (no more than 10 words), 
- Descriptive (capture the essence of the epic), and 
- Convey the primary objective of the epic  (value of
doing the epic).

Example: “Enable Password Reset for Registered Users to 
Regain Account Access."

In your ASSESSMENT please respond ONLY with a valid JSON 
(no other text) in the following format: 

{
   "assessments": [
        {
             "element": "<ELEMENT>",
             "quality": "<QUALITY>",
             "explanation": "<EXPLANATION>"
             "recommendations": "<RECOMMENDATIONS>"
        }
   ]
}

ASSESSMENT:
"""

\end{verbatim}

\end{document}